\documentstyle[aps,pra,twocolumn,epsf]{revtex}

\begin{document}
\draft
\wideabs{
\title{Interatomic collisions in a tightly confined Bose gas} 
\author{D.S. Petrov${}^{1,2}$ and G.V. Shlyapnikov${}^{1,2,3}$}
\address{${}^1$ FOM Institute for Atomic and Molecular Physics, Kruislaan 407,
1098 SJ Amsterdam, The Netherlands \\
 ${}^2$ Russian Research Center,
Kurchatov Institute, 
 Kurchatov Square, 123182 Moscow, Russia \\ ${}^3$
Laboratoire Kastler Brossel,\footnotemark 24 rue Lhomond, F-75231 Paris Cedex
05, France} 
\date{\today}
\maketitle
\begin{abstract}
We discuss pair interatomic collisions in a Bose gas tightly confined in one
(axial) direction and identify two regimes of scattering. In the quasi2D
regime, where the confinement frequency $\omega_0$ greatly exceeds the gas
temperature $T$, the scattering rates exhibit 2D features of the particle
motion. At temperatures $T\sim\hbar\omega_0$ one has a confinement-dominated
3D regime, where the confinement can change the momentum dependence of the 
scattering amplitudes. We describe the collision-induced energy exchange 
between the axial and radial degrees of freedom and analyze recent experiments
on thermalization and spin relaxation rates in a tightly (axially)
confined gas of Cs atoms. 

\end{abstract}
}
%\narrowtext
%\onecolumn
%\narrowtext

\footnotetext{
\footnotemark  LKB is an unit\'{e} de recherche de l'Ecole Normale
Sup\'{e}rieure et de l'Universit\'{e} Pierre et Marie Curie, associ\'{e}e
au CNRS.
}
\section{Introduction} \label{Sec.Introduction}

Collisional properties of ultra-cold gases strongly confined in one direction 
attract a great deal of interest since the start of active studies of 
spin-polarized atomic hydrogen. In the latter case the interest was related 
to recombination and spin relaxation collisions and to elastic
scattering in the (quasi)2D gas of atomic hydrogen adsorbed on liquid He
surface (see \cite{Jook} for review). The discovery of Bose-Einstein
condensation in trapped  alkali-atom clouds
\cite{discoveryJila,discoveryMit,discoveryRice} stimulated a progress in
evaporative and optical cooling and in trapping of neutral atoms. Present
facilities make it possible to (tightly) confine the motion of particles in
one direction to zero point oscillations. Then, kinematically the gas is 2D,
and the only difference from the purely 2D case is related to the value of the
interparticle interaction which now depends on the tight
confinement. Thus, one now has many more opportunities to create
(quasi)2D gases.  In the recent experiments with optically trapped Cs 
\cite{Chu1,Chu2,Christ1,Christ2} about 90$\%$ of atoms are accumulated in the 
ground state of the harmonic oscillator potential in the direction of the tight
confinement.  

In this paper we consider a Bose gas tightly confined 
in one (axial) direction and discuss how the axial confinement
manifests itself in pair elastic and inelastic collisions. We identify two
regimes of scattering. At temperatures $T\ll\hbar\omega_0$ ($\omega_0$ is 
the axial frequency) only the ground state of the axial harmonic oscillator 
is occupied, and one has a quasi2D regime. In this case, the 2D
character of the relative motion of particles at large separation between them,
manifests itself in a logarithmic energy  dependence of the scattering
amplitude. For a negative 3D scattering length $a$, we observe resonances in
the dependence of the elastic scattering rate on $a$. This is quite different
from the 3D case where the scattering rate always increases with $a^2$. The 
presence of these resonances in quasi2D follows from the 
analysis given in \cite{PHS} and finds its origin in increasing role of 
the 2D kinematics of the particle 
motion with increasing ratio $|a|/l_0$, 
where $l_0=(\hbar/m\omega_0)^{1/2}$ is the
axial extension of the atom wavefunction, and $m$ the atom mass. 

At temperatures $T\sim\hbar\omega_0$
we have a confinement-dominated 3D regime of scattering, where the 2D
character of the particle motion is no longer pronounced in the scattering
process, but the axial confinement can strongly influence the energy
(temperature) dependence of the scattering rate.  Treating collisions as
three-dimensional, the wavevector $p$ of the relative motion of colliding atoms
does not decrease with $T$. The atoms undergo zero-point oscillations in the
axial direction and this corresponds to $p\sim 1/l_0$. If the 3D scattering
amplitude  is momentum-dependent at these $p$, which is the case for $|a|\agt
l_0$, then the temperature dependence of the elastic collisional rate becomes
much weaker. This means that for a large 3D scattering length the tight axial
confinement suppresses a resonant enhancement of the collisional rate at low
energies. In many of the current experiments with
ultra-cold gases one tunes $a$ to large positive or negative values by varying
the magnetic field and achieving Feshbach resonances
\cite{Verhaar,FeshbachKet,FeshbachHeinzen,FeshbachChu,FeshbachJila}. 
In the unitarity
 limit ($|a|\rightarrow \infty$) the 3D elastic 
cross-section is $\sigma=8\pi/p^2$
and the rate of 3D elastic collisions strongly increases with decreasing
temperature. The tight confinement of the axial motion makes the scattering
rate practically temperature independent at $T\sim\hbar\omega_0$.  We obtain a
similar suppression of resonances for  inelastic collisions, 
where the resonant
temperature dependence in 3D is related to the energy dependence of the
initial wavefunction of colliding atoms. We analyze the Stanford and ENS
experiments on elastic \cite{Chu2,Christ2} and spin relaxation \cite{Chu2}
collisions in a tightly axially confined gas of cesium atoms and discuss the
origin of significant deviations of the observed collisional rates from the 3D
behavior. 

We develop a theory to describe the collision-induced energy exchange between
axial and radial degrees of freedom of the particle motion. We establish
selection rules for transitions between particle states in the axial harmonic
potential and calculate the corresponding transition amplitudes. This allows
us to consider temperatures $T\agt\hbar\omega_0$ and analyze
thermalization rates in non-equilibrium clouds. In the Stanford and ENS
experiments these clouds were created by means of degenerate Raman
sideband cooling \cite{Chu1,Chu2,Christ1,Christ2} which effectively leads to a
gas with different axial ($T_z$) and radial ($T_\rho$) temperatures. After
the cooling is switched off, the temperatures $T_z$ and $T_\rho$ 
start to approach
each other, and ultimately the gas reaches the equilibrium temperature. At
sufficiently low $T$ only a few axial states are occupied and the temperature
dependence of the corresponding thermalization rates should deviate from the
3D behavior, thus exhibiting the influence of the axial confinement on the
scattering process. We calculate the thermalization rates and establish the
conditions under which this influence is pronounced. 

The minimum energy exchange between the radial and axial degrees of freedom
of two colliding atoms is equal to $2\hbar\omega_0$. This follows from
the symmetry of the interatomic potential with respect to simultaneous
inversion
 of the axial coordinates of the two atoms, which ensures the
conservation of 
 parity of their wavefunction under this operation.
Accordingly, the sum of 
 two (axial) vibrational quantum
numbers can be changed only by an even value. The rate of energy transfer from
the radial to axial motion is proportional to the difference between the
radial and axial temperatures $\Delta T=T_\rho -T_z$, if they are close to
each other. As the total energy of colliding particles
should exceed $2\hbar\omega_0$ in order to enable the energy transfer, the
rate of this process at temperatures $T<\hbar\omega_0$ 
becomes exponentially small: $\Delta\dot E\propto
\Delta T\exp(-2\hbar\omega_0/T)$.  Due to the presence of the energy
gap
 $\hbar\omega_0$ in the excitation spectrum of the axial harmonic
oscillator, the heat capacity of the axial degree of freedom is
$dE_z/dT_z\sim\exp(-\hbar\omega_0/T)$, which leads to a thermalization rate
$\Delta\dot T/\Delta T\propto\exp(-\hbar\omega_0/T)$. This exponential
temperature dependence shows that the thermalization is suppressed at very low
temperatures. One can deeply cool the axial motion, but radially the cloud
remains "hot" on a very long time scale.

\section{2D scattering problem} \label{Sec.Pure}

First, we discuss the purely 2D elastic scattering in pair collisions of
ultra-cold atoms interacting via a short-range potential $U(\rho)$. At
interparticle distances $\rho\rightarrow\infty$ the wavefunction of colliding
atoms is represented as a superposition of the incident plane wave and scattered
circular wave \cite{LLQ}: 
\begin{equation} \label{PurAsym}  
\psi(\mbox{\boldmath$\rho$})\approx e^{i{\bf
q}\mbox{\boldmath$\rho$}}-f(q,\phi)\sqrt{\frac{i}{8\pi q\rho}}e^{iq\rho}. 
\end{equation}  
The quantity $f(q,\phi)$ is the scattering amplitude, $q$ is the relative
momentum of the atoms, and $\phi$ the scattering angle. Note that $f(q,\phi)$
in Eq.(\ref{PurAsym}) differs by a factor of $-\sqrt{8\pi q}$ from the 2D
scattering amplitude defined in \cite{LLQ}.

Similarly to the 3D case, the scattering amplitude is governed by the
contribution of the $s$-wave scattering if the relative momentum $q$ satisfies
the inequality $qR_e\ll 1$, where $R_e$ is the characteristic radius of
interaction. In the case of alkali atoms, the radius $R_e$ is determined by
the Van der Waals tail of the potential $U(\rho)$ and ranges from 20 \AA$\,$
for Li to 100 \AA$\,$ for Cs. The $s$-wave scattering amplitude is independent
of the scattering angle $\phi$. The probability $\alpha(q)$ for a scattered
particle to pass through a circle of radius $\rho$ per unit time is equal to
the intensity of the scattered wave multiplied by $2\pi\rho v$, where
$v=2\hbar q/m$ is the relative velocity of colliding atoms. From
Eq.(\ref{PurAsym}) we have
\begin{equation}     \label{Puralpha}
\alpha(q)=\frac{\hbar}{2m}|f(q)|^2.
\end{equation}
The velocity $v$ is equal to the current density in the incident wave of
Eq.(\ref{PurAsym}). The ratio of $\alpha(q)$ to this quantity is the 2D cross
section which has the dimension of length: 
\begin{equation} \label{Pursigma}
\sigma(q)=|f(q)|^2/4q.
\end{equation}
For the case of identical bosons Eqs.~(\ref{Puralpha}) and (\ref{Pursigma}) 
have an extra factor 2 in the rhs.

The quantity $\alpha(q)$ is nothing else than the rate constant of elastic
collisions at a given $q$. The average of $\alpha(q)$ over the momentum
distribution of atoms, multiplied by the number of pairs of atoms in a
unit volume, gives the number of scattering events in this volume per unit
time.

For finding the $s$-wave scattering amplitude one has to solve the
Schr\"odinger equation for the $s$-wave of the relative motion of
colliding atoms at energy $\varepsilon=\hbar^2q^2/m$: 
\begin{equation}    
\label{Pur2DSchr} 
\left[-\frac{\hbar^2}{m}\Delta_{\rho}+U(\rho)\right]
\psi_s(q,\rho)=\frac{\hbar^2q^2}{m}\psi_s(q,\rho).    \end{equation}   
At distances $\rho\gg R_e$ the relative
motion is free and one can omit the interaction between atoms. Then the
solution of Eq.(\ref{Pur2DSchr}), which for $q\rho\gg 1$ gives the partial
$s$-wave of $\psi(\mbox{\boldmath$\rho$})$ (\ref{PurAsym}), takes the form
\begin{equation} \label{s-wave}
\psi_s(q,\rho)=J_0(q\rho)-\frac{if(q)}{4}H_0(q\rho),\,\,\,\,\,
\rho\ll R_e,
\end{equation}
where $J_0$ and $H_0$ are the Bessel and Hankel functions.

On the other hand, at distances $\rho\ll 1/q$ one can omit the relative
energy of particles in Eq.(\ref{Pur2DSchr}). The resulting (zero
energy) solution depends on the momentum $q$ only through a normalization
coefficient. In the interval of distances where $R_e\ll\rho\ll 1/q$, the
motion is free and this solution becomes $\psi_s\propto \ln(\rho/d)$, where 
$d>0$ is a characteristic length that depends on a detailed shape of the 
potential $U(\rho)$
 and has to be found from the exact solution of
Eq.(\ref{Pur2DSchr}) with
 $q=0$. This logarithmic expression serves as a
boundary condition for
 $\psi_s(q,\rho)$ (\ref{s-wave}) at $q\rho\ll 1$, which
immediately leads to the
 scattering amplitude \cite{LLQ}   
\begin{equation}    \label{Purf} 
f(q)=\frac{2\pi}{\ln(1/qd_*)+i\pi/2}, 
\end{equation} 
where $d_*=(d/2)\exp{C}$ and $C\approx 0.577$ is the Euler constant.

It is important to mention that the condition $qR_e\ll 1$ is sufficient for  
the validity of Eq.(\ref{Purf}). This equation also holds for
the case of resonance scattering, where the potential $U(\rho)$ supports a
real (or virtual) weakly bound $s$-level. In this case the spatial shape of
$\psi_s(q,\rho)$ at distances where $R_e\ll\rho\ll 1/q$, is the same as the
shape of the wavefunction of the weakly bound state. This gives 
$d_*=\hbar/\sqrt{m\varepsilon_0}$, where $\varepsilon_0$ is the binding
energy. We thus have the inequality $d_*\gg R_e$, and the quantity $qd_*$ in
Eq.(\ref{Purf}) can be both small and large. 
The rate constant $\alpha(q)$ peaks at $q=1/d_*$ and decreases as 
$1/[1+(4/\pi^2)\ln^2(qd_*)]$ with increasing or decreasing $q$. Note that
the 2D resonance is actually a resonance in the logarithmic scale of energies.
The decrease of $\alpha$ by factor 2 from its maximum value requires a change
of energy $\varepsilon=\hbar^2q^2/m$ by factor 20.

For $qd_*\ll 1$ one may omit the imaginary part in Eq.(\ref{Purf}), and the
scattering amplitude becomes real and positive \cite{com3}. The positive sign
of $f(q)$ has a crucial consequence for the mean-field interparticle
interaction in purely 2D Bose gases. In the ultra-cold limit where $qR_e\ll
1$,  the scattering amplitude is related to the energy of interaction in a
pair  of particles (coupling constant $g$). For a short-range
potential $U(\rho)$, the energy of the mean-field interaction in a weakly
interacting gas is the sum of all pair interactions. In a uniform
Bose-condensed gas the coupling constant $g$ for condensate atoms is equal to
the amplitude of scattering (with an extra factor $\hbar^2/m$ for our
definition of $f$)  at the energy of the relative motion
$\varepsilon=\hbar^2q^2/m=2\mu$, where $\mu$ is the chemical potential
\cite{com1}. Hence, we have   
\begin{equation}     \label{gfin}
g=\frac{\hbar^2}{m}f(q_c)=\frac{2\pi\hbar^2}{m}\frac{1}{\ln(1/q_cd_*)}>0;
\,\,\,\,\, q_cd_*\ll 1,  
\end{equation}     
where $q_c=\sqrt{2m\mu}/\hbar$ is the inverse healing length. In a dilute  
thermal 2D gas, due to the logarithmic dependence of $f$ on $q$, the
thermal average of the mean-field interaction leads to
the coupling constant $g=(\hbar^2/m)f(q_T)$, where $q_T=\sqrt{mT}/\hbar$   is
the thermal momentum of particles. At sufficiently low temperatures,  where
$q_Td_*\ll 1$, we again have $g>0$. 

Thus, in an ultra-cold purely 2D gas the coupling constant for the
mean-field interaction is always positive in the dilute limit
and, hence, the interaction is repulsive. This striking difference from the 3D
case is a consequence of the 2D kinematics. For low energies, at interparticle
distances $\rho\gg R_e$, the (free) relative motion of a pair of atoms is
governed by the wavefunction $\psi_s\propto\ln{(\rho/d)}$. The probability
density $|\psi_s|^2$ of finding two atoms at a given separation increases with
$\rho$ as the condition $\rho>d$ is always reached, unless the atoms have  a
bound state with energy $\varepsilon\rightarrow 0$ ($d\rightarrow\infty$). This
means that it is favorable for particles to be at larger $\rho$, i.e. they
repel each other.

\section{Scattering in axially confined geometries. 
General approach} \label{Sec.Amplitude}

In this Section we discuss elastic scattering of atoms (tightly) confined
in the axial ($z$) direction, assuming that the motion in two other
($x,y$) directions is free. We analyze how the scattering is influenced by the
confinement and calculate a complete set of scattering amplitudes corresponding
to collision-induced transitions between particle states in the 
confining potential. We still call this scattering elastic as the
internal states of atoms are not changing.

For a harmonic axial confinement, the motion of two atoms interacting 
with  each other via the potential $V(r)$
can be still separated into their relative and  center-of-mass motion. 
The latter drops out of the scattering problem.
The relative motion is governed by the potential $V(r)$, together with the 
potential $V_H(z)=\omega_0^2z^2/4$ originating from the axial 
confinement with frequency $\omega_0$. For the incident wave 
characterized by the wavevector ${\bf q}$ 
of the motion in the $x,y$ plane and by the quantum number $\nu$ of the
state in the potential $V_H(z)$, the wavefunction of the relative motion 
satisfies the Schr\"odinger equation   
\begin{equation}     \label{Schr}  
\left[-\frac{\hbar^2}{m}\Delta+V(r)+V_H(z)-\frac{\hbar\omega_0}{2}\right]
\psi({\bf r})=\varepsilon\psi({\bf r}),   \end{equation}   
where $\varepsilon=\hbar^2q^2/m+\nu\hbar\omega_0$. 
 
The scattering depends crucially on the relation between the
radius of interatomic interaction $R_e$ and the characteristic de Broglie 
wavelength of particles $\tilde\Lambda_{\varepsilon}$. The latter is 
introduced qualitatively, as the motion along the $z$ axis is tightly 
confined. Accounting for the zero point axial oscillations one can write
$\tilde\Lambda_{\varepsilon}\sim\hbar/\sqrt{m\tilde\varepsilon}$, with
$\tilde\varepsilon=\varepsilon+\hbar\omega_0/2$.   
We will consider the ultra-cold limit where 
\begin{equation}    \label{u}
\tilde\Lambda_{\varepsilon}\gg R_e.
\end{equation}
Eq.(\ref{u}) immediately leads to the inequality
$qR_e\ll 1$,
as the de Broglie wavelength for the motion in the $x,y$ plane is $\sim 1/q$.
For small $\nu$ the harmonic oscillator length $l_0=(\hbar/m\omega_0)^{1/2}$
plays the role of the axial de Broglie wavelength of atoms. Therefore, the 
ultra-cold limit (\ref{u}) also requires the condition $l_0\gg R_e$. For 
large $\nu$, the axial de Broglie wavelength is 
$\sim l_0/\sqrt{\nu}$ and, according to Eq.(\ref{u}), this quantity should 
be much larger than $R_e$.  

Under the condition $qR_e\ll 1$, the scattering amplitudes are determined 
by the contribution of the
$s$-wave for the motion in the $x,y$ plane. In the case of identical bosons,
the $s$-wave scattering requires even values of $\nu$ and $\nu'$ as the
wavefunction $\psi$ should conserve its sign under the transformation
$z\rightarrow -z$. The
quantum numbers $\nu$ and $\nu'$ should be even also for distinguishable
particles. Otherwise at distances of interatomic interaction, $r\alt R_e$, the
wavefunction $\psi$ will be small at least as $R_e/l_0$, ensuring the 
presence of this small parameter in the expressions for the 
scattering amplitudes.

The scattering amplitudes corresponding to transitions from the initial state
$\nu$ (of the relative motion in the potential $V_H(z)$) to final states
$\nu'$ are defined through the asymptotic form of the wavefunction $\psi$ at an
infinite separation $\rho$ in the $x,y$ plane:
\begin{equation} \label{Asym}  
\!\psi({\bf r})\!\approx\!\varphi_{\nu}(z)e^{i{\bf
q}\mbox{\boldmath$\rho$}}\!-\!\sum_{\nu'}f_{\nu\nu'}(\varepsilon)\varphi_{\nu'}(z)
\sqrt{\frac{i}{8\pi q_{\nu'}\rho}}e^{iq_{\nu'}\rho}, 
\end{equation} 
where $\varphi_{\nu}(z)$ and $\varphi_{\nu'}(z)$ are the (real) eigenfunctions
of the states $\nu$ and $\nu'$. For each of the scattered circular waves the
value of the momentum $q_{\nu'}$ follows from the energy conservation law
$\hbar^2q_{\nu'}^2/m=\varepsilon-\hbar\omega_0\nu'>0$.

Relying on the condition (\ref{u}), we develop a method 
that allows us to express the scattering
amplitudes through the 3D scattering length.
At interparticle distances $r\gg R_e$ the relative motion in the $x,y$ plane
is free, and the motion along the $z$ axis is governed only by the harmonic
oscillator potential $V_H(z)$. Then, the solution of Eq.(\ref{Schr}) with
$V(r)=0$ can be expressed through the Green function $G_{\varepsilon}
({\bf r},{\bf r}')$ of this equation. Retaining only the $s$-wave for 
the motion in the $x,y$ plane, we have 
\begin{equation}    \label{Green}  
\psi({\bf r})=\varphi_{\nu}(z)J_0(q\rho)+A_{\nu}G_{\varepsilon}({\bf r},0),  
\end{equation}  
and the expression for the Green function $G_{\varepsilon}({\bf r},0)$ reads
\renewcommand{\arraystretch}{1.5}  
\begin{equation}    \label{EG} 
\!\!G_{\varepsilon}({\bf
r},0)\!=\!\!\sum_{\nu'}\!\varphi_{\nu'}(z)\varphi_{\nu'}(0)\!\!\times\!\!
\left\{\begin{array}{l}\!\!iH_0^{(1)}(q_{\nu'}\rho)/4;\,q^2_{\nu'}\!>\!0
\\\!\!K_0(|q_{\nu'}|\rho)/2\pi;\,q^2_{\nu'}\!<\!0\end{array}\right.
\end{equation}  
\renewcommand{\arraystretch}{1}
\hspace{-1.5mm}Here the summation is also performed over closed scattering
channels for which $q_{\nu'}^2<0$. The function $K_0(x)=(i\pi /2)H_0(ix)$
and it decays as $\sqrt{\pi/2x}\,\exp{(-x)}$ at $x\gg 1$. Thus, for 
$\rho\rightarrow\infty$ the terms corresponding to the closed channels vanish.
Then, comparing Eq.(\ref{Green}) at $\rho\rightarrow\infty$ with 
Eq.(\ref{Asym}), we find a
relation between the scattering amplitudes and the coefficients $A_{\nu}$:
\begin{equation} \label{Ampl} 
f_{\nu\nu'}=-A_{\nu}\varphi_{\nu'}(0)\theta(\varepsilon-\hbar\omega_0\nu'), 
\end{equation} 
where $\theta$ is the step function.

The condition $l_0\gg R_e$ ensures that the relative motion of atoms in the
region of interatomic interaction is not influenced by the axial (tight)
confinement. Therefore, the wavefunction $\psi({\bf r})$ in the interval of
distances where  $R_e\ll r\ll\tilde\Lambda_{\varepsilon}$,
differs only by a normalization coefficient from the 3D wavefunction of free
motion at zero energy. Writing this coefficient as $\varphi_{\nu}(0)\eta$, we
have 
\begin{equation}   \label{3D}  
\psi(r)\approx\varphi_{\nu}(0)\eta (1-a/r).   
\end{equation} 
Eq.(\ref{3D}) serves as a boundary condition for $\psi({\bf r})$ (\ref{Green})
at $r\rightarrow 0$. 

For $r\rightarrow 0$, a straightforward calculation of the sum in Eq.(\ref{EG})
yields
\begin{equation} \label{lim}
G_{\varepsilon}(r,0)\approx\frac{1}{4\pi
r}+\frac{1}{2(2\pi)^{3/2}l_0} w\left(\frac{\varepsilon}{2\hbar\omega_0}\right),
\end{equation}
where the complex function $w(x)$ is given by
\begin{equation} \label{w}
\!\!w(x\!)\!\!=\!\!\lim_{\!N\!\rightarrow\!\infty}\!\!\left[\!2\sqrt{\frac{N}{\pi}}\ln\frac{N}{e^2}
\!-\!\!\sum_{j=0}^N\!\frac{(2j\!-\!1)!!}{(2j)!!}\ln(j\!-\!x\!-\!i0)\!\right]\!. \!\!
\end{equation}  
With the Green function (\ref{lim}), the wavefunction (\ref{Green}) at 
$r\rightarrow 0$ should coincide with $\psi(r)$ (\ref{3D}). This gives the 
coefficient 
\begin{equation}   \label{eta}
\eta=\frac{1}{1+(a/\sqrt{2\pi}l_0)w(\varepsilon/2\hbar\omega_0)} 
\end{equation}
and provides us with the values of the coefficients $A_{\nu}$. Then, using
Eq.(\ref{Ampl}) and explicit expressions $\varphi_{\nu}(0)=(1/2\pi l_0^2)^{1/4}
\!(\nu-1)!!/\sqrt{\nu!}$, we
immediately obtain the scattering amplitude $f_{00}(\varepsilon)$ and express
all other scattering amplitudes through this quantity: 
\begin{equation}    \label{Amplfin} 
f_{00}(\varepsilon)=4\pi\varphi_0^2(0)a\eta=\frac{2\sqrt{2\pi}}
{l_0/a+(1/\sqrt{2\pi})w(\varepsilon/2\hbar\omega_0)},   
\end{equation}
\begin{equation}   \label{Amplfinnu}
f_{\nu\nu'}(\varepsilon)=P_{\nu\nu'}f_{00}(\varepsilon)
\theta(\varepsilon-\hbar\omega_0\nu)\theta(\varepsilon-\hbar\omega_0\nu'), 
\end{equation}
where
\begin{equation}   \label{P}
P_{\nu\nu'}=\frac{\varphi_{\nu}(0)\varphi_{\nu'}(0)}{\varphi_0^2(0)}=
\frac{(\nu-1)!!(\nu'-1)!!}{\sqrt{\nu!\nu'!}}.
\end{equation}

One can see from Eqs.~(\ref{Amplfin}) and (\ref{Amplfinnu}) that for any
transition $\nu\rightarrow\nu'$ the scattering amplitude is a
universal function of the parameters $a/l_0$ and $\varepsilon/\hbar\omega_0$.
The quantity $P_{\nu\nu'}$ in Eq.(\ref{Amplfinnu}) is nothing else than the 
relative probability amplitude of having an axial interparticle separation
$|z|\ll l_0$ (in particular, $|z|\alt R_e$) for both incoming ($\nu$) and
outgoing ($\nu'$) channels of the scattering process. It is thus sufficient 
to study only the behavior of $f_{00}(\varepsilon)$. 

We emphasize the presence of two distinct regimes of scattering. The first
one, which we call quasi2D, requires relative energies 
$\varepsilon\ll\hbar\omega_0$. In this case, the relative motion of particles 
is confined to zero point oscillations in the axial direction, and the 2D
kinematics of the relative motion at interatomic distances $\rho>l_0$ should
manifest itself in the dependence of the scattering amplitude on 
$\varepsilon/2\hbar\omega_0$ and $a/l_0$. In the other regime, at energies
already comparable with $\hbar\omega_0$, the 2D kinematics is no longer
pronounced in the scattering process. Nevertheless, the latter is still
influenced by the (tight) axial confinement. Qualitatively, the scattering
amplitudes become three-dimensional, with a momentum 
$\sim 1/l_0$ related to the quantum character
of the axial motion.  Thus, we can say that this is a confinement-dominated 
3D regime of scattering. With increasing the relative energy to 
$\varepsilon\gg\hbar\omega_0$, the momentum is increasing to 
$\sqrt{m\varepsilon}/\hbar$ and the confinement-dominated 3D regime 
continuously transforms to ordinary 3D scattering. 

\section{Quasi2D regime} \label{Sec.Quasi2D} 

In the quasi2D regime, due to the condition $\varepsilon\ll\hbar\omega_0$,
the incident and scattered waves have quantum numbers
$\nu=\nu'=0$ for the motion in the axial harmonic potential $V_H(z)$. The
relative energy $\varepsilon=\hbar^2q^2/m$ and the inequality $ql_0\ll 1$
is satisfied. In this case Eq.(\ref{w}) gives 
\begin{equation}    \label{wlog}
w(\varepsilon/2\hbar\omega_0)=\ln{(B\hbar\omega_0/\pi\varepsilon)}+i\pi, 
\end{equation}
where $B\approx 0.915$. Then our equation (\ref{Amplfin}) recovers Eq.(11) 
of ref. \cite{PHS}, obtained in this limit \cite{com2}. 

Using Eq.(\ref{wlog}) we can represent $f_{00}(\varepsilon)$ (\ref{Amplfin})
in the 2D form (\ref{Purf}), with
\begin{equation}   \label{dquasi}
d_{*}=(d/2)\exp{C}=\sqrt{\pi/B}\,l_0\exp{(-\sqrt{\pi/2}\,l_0/a)}.
\end{equation}
This fact has a physical explanation. Relying on the same
arguments as in the purely  2D case, one finds that in the interval of
distances where $l_0\ll\rho\ll 1/q$, the wavefunction
$\psi\propto\varphi_0(z)\ln{(\rho/d)}$. On the other hand, for $\rho\gg l_0$ 
we have  $\psi({\bf r})=\varphi_0(z)\psi_s(\rho)$,  where $\psi_s$ is given by
the  2D expression (\ref{s-wave}) with $f(q)=f_{00}(\varepsilon)$. This follows
from Eqs.~(\ref{Green})-(\ref{Ampl}), as all closed scattering channels
($\nu'\neq 0$) in Eq.(\ref{EG}) for the Green function 
$G_{\varepsilon}({\bf r},0)$ have momenta $|q_{\nu'}|\agt 1/l_0$ and will be
exponentially suppressed at $\rho\gg l_0$. Matching the two expressions for
the wavefunction $\psi$ one immediately obtains the 2D equation (\ref{Purf}).
However, the parameter $d_*$ (\ref{dquasi}) can be found only from the
solution of the quasi2D scattering problem.  

We thus conclude that the scattering problem in the quasi2D regime is 
equivalent to the scattering in an effective purely 2D potential which leads
to the same value of $d_*$. For positive $a\ll l_0$, this potential can 
be viewed as a (low) barrier, 
with a height $V_0\sim \hbar^2a/ml_0^3$ and radius $l_0$.
Hence, in the case of positive $a$ we have a small (positive) scattering 
amplitude, in accordance with Eqs.~(\ref{Purf}) and (\ref{dquasi}).
For a negative $a$ satisfying the condition $|a|\ll l_0$, the effective
potential is a shallow well which has a depth $|V_0|$ and radius $l_0$. This
shallow well supports a weakly bound state with an exponentially small binding
energy $\varepsilon_0$, which leads to an exponentially large $d_*$ 
as follows from Eq.(\ref{dquasi}). As a result, we have a resonance energy 
dependence of the scattering amplitude $f_{00}$
at a fixed ratio $a/l_0$, and a resonance behavior of $f_{00}$ as a function
of $a/l_0$ at a fixed $\varepsilon/\hbar\omega_0$. 

The resonance in the energy dependence of $f_{00}$ is quite similar to the
logarithmic-scale resonance in the purely 2D case, discussed in Section 
\ref{Sec.Pure}. The quasi2D resonance is also described by Eq.(\ref{Purf}),
where the length $d_*$ is now given by Eq.(\ref{dquasi}). As expected,
the dependence of $f_{00}$ on $\varepsilon$ is
smooth.

The resonance in the dependence of the quasi2D scattering amplitude on 
$a/l_0$ has been found and discussed in \cite{PHS}. Relying on the above
introduced effective 2D potential for the quasi2D scattering, we can 
now explain this resonance on the same grounds as the 
resonance in the energy dependence of $f_{00}$. We will do this in terms
of the relative energy $\varepsilon$ and the binding energy
in the effective potential, 
$\varepsilon_0=\hbar^2/md_*^2\propto\exp{(l_0/|a|)}$.
For $\varepsilon/\varepsilon_0=(qd_*)^2\gg 1$, 
the scattering amplitude in Eq.(\ref{Purf}) is real and negative. 
It increases in magnitude with decreasing ratio 
$\varepsilon/\varepsilon_0$, that is with decreasing $q$ or $l_0$. 
In the opposite limit, where $\varepsilon/\varepsilon_0\ll 1$, the scattering
amplitude is real and positive and it increases with the ratio 
$\varepsilon/\varepsilon_0$. The region of energies 
$\varepsilon/\varepsilon_0\sim 1$ corresponds to the resonance, where both 
the real and imaginary parts of $f_{00}$ are important. The real part 
reaches its maximum at $\varepsilon/\varepsilon_0=\exp{(-\pi)}$, drops
to zero at $\varepsilon/\varepsilon_0=1$, and acquires the maximum negative 
value for $\varepsilon/\varepsilon_0=\exp{\pi}$. The dependence of
${\rm Im}f_{00}$ on $\varepsilon/\varepsilon_{0}$ is the same as that of
the quantity $|f_{00}|^2$. Both of them peak at $\varepsilon/\varepsilon_0=1$
and decrease with increasing or decreasing $\varepsilon/\varepsilon_0$.

Qualitatively, the picture remains the same for $|a|\sim l_0$.
In Fig.1 we present the dependence of $|f_{00}|^2$ on 
$\varepsilon/2\hbar\omega_0$ at $a/l_0$ equal to $-1$, $1$, and $\infty$. 
In the two last cases we always have $\varepsilon/\varepsilon_0\ll 1$, and 
$|f_{00}|^2$ increases with $\varepsilon$ at $\varepsilon\ll\hbar\omega_0$.
For $a/l_0=-1$ we have the above described logarithmic-scale resonance in 
the behavior of $|f_{00}|^2$.  

\begin{figure} 
\epsfxsize=\hsize 
\epsfbox{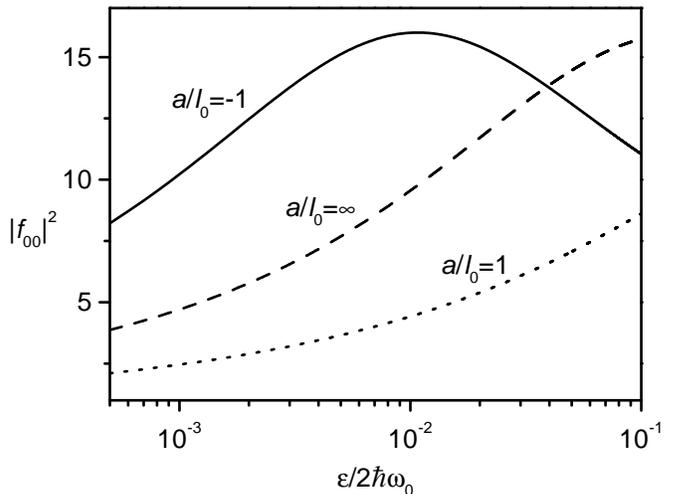} 
\caption{\protect
The function $|f_{00}|^2$ versus energy for $a/l_0=-1$ (solid curve), 
$a/l_0=\infty$ (dashed curve) and $a/l_0=1$ (dotted curve).}    
\label{1}   
\end{figure} 

The quasi2D resonance is much more pronounced in the dependence
of the scattering amplitude on the parameter $a/l_0$. The reason is that
$f_{00}$ logarithmically depends on the particle energy, whereas the dependence
on $a/l_0$ is a power law. For $\varepsilon\ll\hbar\omega_0$,
Eq.(\ref{Amplfin}) yields
\begin{equation}   \label{f2}
|f_{00}|^2=\frac{16\pi^2}{(\sqrt{2\pi}\,l_0/a+
\ln{(B\hbar\omega_0/\pi\varepsilon)})^2+\pi^2}. 
\end{equation}
The quantity $|f_{00}|^2$ differs only by a factor of $\hbar/m$ from the rate
constant of elastic collisions (see Eq.(\ref{Puralpha})), and one can think of
observing the resonance dependence of $|f_{00}|^2$ on  $a/l_0$ in an
experiment. For example, one can keep $\varepsilon$  (temperature) and
$\omega_0$ constant and vary $a$ by using Feshbach resonances. The resonance
is
 achieved at $a=-l_0\ln{(B\hbar\omega_0/\pi\varepsilon)}$. This is a
striking
 difference from the purely 3D case, where the cross section and rate
constant
 of elastic collisions monotonously  increase with $a^2$. 
 
In Fig.2 we present $|f_{00}|^2$ versus $a/l_0$ at a fixed 
$\varepsilon/\hbar\omega_0$. 
In order to extend the results to the region of energies where the 
validity of
the quasi2D approach is questionable, the quantity $|f_{00}|^2$ was 
calculated by using Eq.(\ref{Amplfin}) for the scattering amplitude.
The resonance is still visible at $\varepsilon/\hbar\omega_0=0.06$ and it
disappears for $\varepsilon/\hbar\omega_0=0.2$. 

The obtained results allow us to conclude 
that for $|a|\agt l_0$ the approximate border line between the 
quasi2D and confinement-dominated 
3D regimes is $\varepsilon\approx\varepsilon_*=0.1\hbar\omega_0$.
For $|a|\ll l_0$, as we will see below, the confinement-dominated
regime is practically absent.

\begin{figure} 
\epsfxsize=\hsize 
\epsfbox{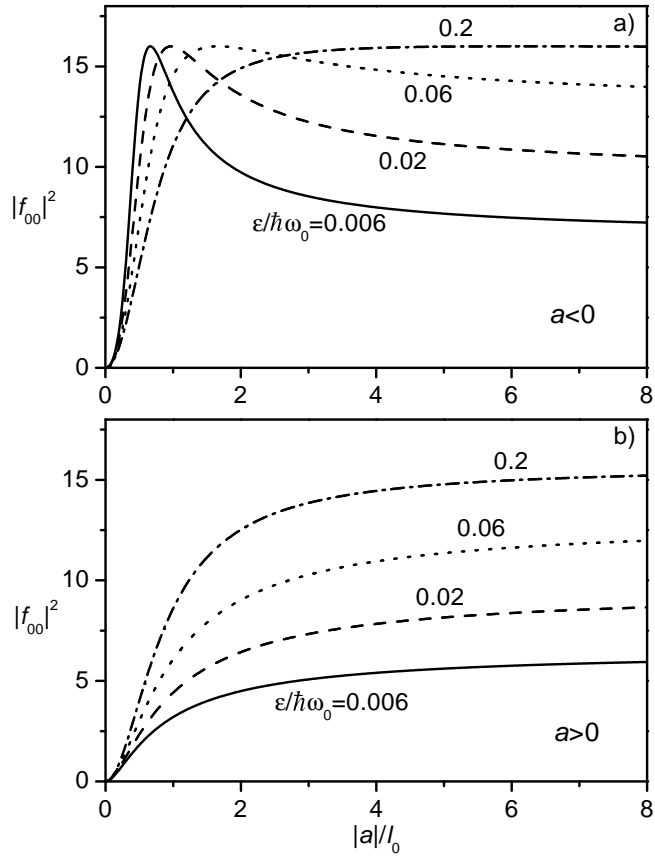} 
\caption{\protect
The function $|f_{00}|^2$ versus $|a|/l_0$ at various energies for $a<0$ (a)
and $a>0$ (b).}   
\label{2}   \end{figure} 

The output of kinetic studies in thermal gases is usually related to the
mean collisional frequency (the rate of interatomic collisions) 
$\Omega=\bar\alpha n$, where $\bar\alpha$ is the mean rate constant of elastic
collisions, and $n$ the gas density. In the quasi2D regime, the rate
constant  $\bar\alpha$ follows directly from Eq.(\ref{Puralpha}),
with twice as large rhs for identical bosons: 
\begin{equation}  \label{alphaquasi}
\bar\alpha=\frac{\hbar}{m}\langle |f_{00}|^2\rangle, 
\end{equation}
where the symbol $\langle\,\rangle$ stands for
the thermal average. Our numerical calculations 
show that the average of $|f_{00}|^2$ over the Boltzmann
distribution of particles only slightly broadens the resonances in Fig.1 
and Fig.2. Due to the logarithmic dependence of
$f_{00}$ on the relative energy, the thermal average is obtained with a
good accuracy if one simply replaces $\varepsilon$ by the gas temperature $T$.
Thus, in order to observe the manifestation of the 2D features of the 
particle motion in their collisional rates one has to achieve very 
low temperatures $T<0.1\hbar\omega_0$.

\section{Confinement-dominated 3D regime}  \label{Sec.Conf-dom}

In the confinement-dominated 3D regime, where $\varepsilon\sim\hbar\omega_0$,
the axial confinement influences the scattering process through the confined 
character of the axial motion. In order to analyze this influence, we first 
examine the function 
$w(\varepsilon/2\hbar\omega_0)$ which determines the energy dependence of 
the scattering amplitudes. The imaginary part of $w(x)$, following from 
Eq.(\ref{w}), is equal to  
\begin{equation} \label{Im}
{\rm Im} w(x)=\pi\sum_{j=0}^{[x]}\frac{(2j-1)!!}{(2j)!!}=2\sqrt{\pi}
\frac{\Gamma([x]+3/2)}{[x]!},
\end{equation}
where $[x]$ is the integer part of $x$. The function ${\rm Im} w(x)$ has a
step-wise behavior as shown in Fig.3. It is  constant at non-integer $x$
and undergoes a jump at each integer $x$,  taking a larger value for larger
$x$. With increasing $x$, the jumps become smaller and for $x\gg 1$ we have
${\rm Im}w(x)\approx 2\sqrt{\pi x}$.  The real part of $w(x)$ was calculated
numerically from Eq.(\ref{w}) and is also given in Fig.3. At any $x$ we have
$|{\rm Re}w(x)|<1$, except for narrow intervals in the vicinity of
integer $x$.  In each of these intervals the function ${\rm Re}w(x)$ 
logarithmically goes to infinity as $x$ approaches the corresponding integer
value. This is consistent with the step-wise behavior of ${\rm Im}w(x)$: As
one can see directly from  Eq.(\ref{w}), for $x$ approaching an integer $j$
the analytical complex function $w\propto\ln{(j-x-i0)}$. 

\begin{figure} 
\epsfxsize=\hsize 
\epsfbox{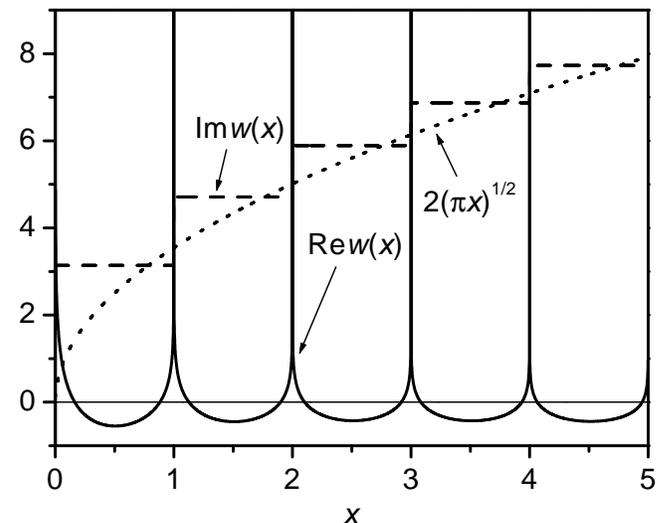} 
\caption{\protect
The functions ${\rm Re}w(x)$ (solid curve) and ${\rm Im}w(x)$ (dashed lines).
The dotted curve shows the function $2\sqrt{\pi x}$ corresponding to the
asymptotic behavior of ${\rm Im}w$ at large $x$.}    
\label{3}   
\end{figure}

The described behavior of the function $w(\varepsilon/2\hbar\omega_0)$
has a direct influence on the scattering amplitudes.
For $\varepsilon/2\hbar\omega_0$ close to an integer $j$, the amplitude is 
small and it is equal to zero for $\varepsilon=2\hbar\omega_0j$. This 
phenomenon originates from the fact that for $\varepsilon$ close to 
$2\hbar\omega_0j$, a new scattering channel opens (really or virtually). 
For this channel the momentum 
$|q_{\nu'}|=\sqrt{m|\varepsilon-2\hbar\omega_0j|}/\hbar$ is very small.
Hence, at distances $\rho\ll |1/q_{\nu'}|$ the wavefunction $\psi$ (\ref{Green})
will be determined by the contribution of this low-momentum term if 
$\rho\gg l_0$. This is clearly seen from Eqs.~(\ref{EG}) and (\ref{Green}) 
and makes the situation somewhat similar to that in the quasi2D regime of 
scattering. In the latter case, the wavefunction $\psi$ (\ref{Green}) at 
distances $\rho\ll 1/q$ is also determined by the contribution of the 
low-momentum channel as long as $\rho\gg l_0$. Then, as follows from the 
analysis in Section \ref{Sec.Quasi2D},
this wavefunction and the scattering amplitude $f_{00}$ behave as 
$1/\ln{(\hbar\omega_0/\varepsilon)}$ in
the limit $\varepsilon\rightarrow 0$.
In the present case, the wavefunction $\psi$ and the scattering
amplitudes are small as 
$1/\ln{(\hbar\omega_0/|\varepsilon-2\hbar\omega_0j|)}$ for 
$\varepsilon\rightarrow 2\hbar\omega_0j$.    

\begin{figure} 
\epsfxsize=\hsize 
\epsfbox{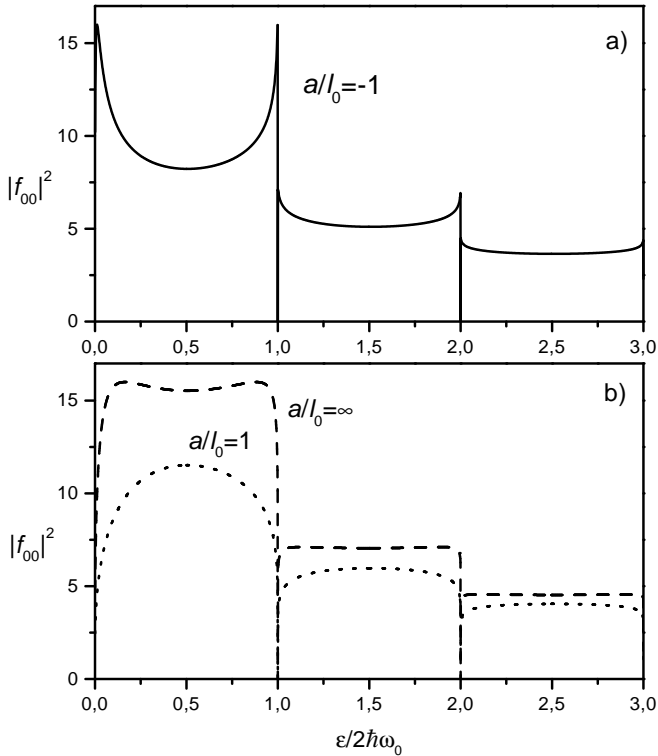} 
\caption{\protect
The function $|f_{00}|^2$ versus energy. In (a) the parameter $a/l_0=-1$. 
In (b) the dashed curve corresponds to the unitarity limit ($a/l_0=\infty$),
and the dotted curve to $a/l_0=1$.}    \label{4}  
\end{figure} 

The energy dependence of $|f_{00}|^2$ on $\varepsilon/2\hbar\omega_0$ for
$a/l_0$ equal to $-1$, $1$, and $\infty$ is displayed in Fig.4. Outside 
narrow energy intervals in the vicinity of integer 
$\varepsilon/2\hbar\omega_0$, the quantity $|f_{00}|^2$ is a smooth function
of $\varepsilon$. One can also see a sort of a step-wise decrease of
$|f_{00}|^2$ with increasing $\varepsilon$, originating from the step-wise
increase of the function ${\rm Im}w(\varepsilon/2\hbar\omega_0)$. For 
$\varepsilon/2\hbar\omega_0$ close to integer values $j>0$ we find a fine 
structure similar in nature to the behavior of $|f_{00}|^2$ at
$\varepsilon\ll\hbar\omega_0$.
For $a/l_0=1$ and $a/l_0=\infty$ there are narrow dips corresponding to the
logarithmic decrease of $|f_{00}|^2$ as 
$\varepsilon\rightarrow 2\hbar\omega_0j$, and for $a/l_0=-1$
these dips are accompanied by resonances. 
Note that the thermal distribution of particles
averages out this fine structure, and the latter will not be pronounced in
kinetic properties.

The difference between the confinement-dominated 3D regime and the ordinary
3D regime of scattering will manifest itself in the rate of 
elastic collisions (mean collisional frequency $\Omega$). For the Boltzmann
distribution of particles, one can find this quantity by turning to the
thermal distribution for the relative motion of colliding partners. 
Collision-induced transitions between the states of the relative motion
in the axial potential $V_H(z)$ are described by the rate constants
\begin{equation}   \label{alphanu}
\alpha_{\nu\nu'}(\varepsilon)=(\hbar/m)|f_{\nu\nu'}(\varepsilon)|^2,
\end{equation}
where the scattering amplitudes $f_{\nu\nu'}$ are given by 
Eqs.~(\ref{Amplfin})
and (\ref{Amplfinnu}), and an extra factor 2 for identical bosons is taken 
into account. The collisional frequency $\Omega=\bar\alpha n$, where $n$ is
the (2D) density, and the mean rate constant of elastic collisions, 
$\bar\alpha$, is obtained by averaging $\alpha_{\nu\nu'}$ (\ref{alphanu}) 
over the thermal distribution of relative energies $\varepsilon$ and by making
the summation over all possible scattering channels. We thus have
\begin{equation}    \label{baralpha}
\Omega=\bar\alpha n=\sum_{\nu\nu'}\int \frac{n\Lambda_T^2d^2q}{(2\pi)^2}
\alpha_{\nu\nu'}(\varepsilon)A\exp{\left( -\frac{\varepsilon}{T}\right) }.
\end{equation} 
Here $\Lambda_T=(2\pi\hbar^2/mT)^{1/2}$ is the thermal de Broglie wavelength,
$\varepsilon=\hbar^2q^2/m+\hbar\omega_0\nu$, and the quantum 
numbers $\nu$ and $\nu'$ take only even values. The distribution function 
over $\nu$ and $q$ is normalized to unity, and the normalization coefficient 
$A$ is equal to 
\begin{equation}   \label{A}
A=2(1-\exp{(-\hbar\omega_0/T)}).
\end{equation}
Note that the number of collisions per unit time and unit
surface area in the $x,y$ plane is equal to 
$\bar\alpha n^2/2=\Omega n/2$.

The manifestation of the tight axial confinement of the particle motion 
in collisional rates depends on the relation between the scattering length 
$a$ and the characteristic de Broglie wavelength 
$\tilde\Lambda_{\varepsilon}\sim\hbar/\sqrt{m(\varepsilon+\hbar\omega_0/2)}$
accounting for the zero point axial oscillations. For the scattering
length satisfying the condition $|a|\ll\tilde\Lambda_{\varepsilon}$, 
the scattering amplitudes are energy
independent at any $\varepsilon$, except for extremely small energies in
the quasi2D regime. This follows directly from 
Eqs.~(\ref{eta})-(\ref{Amplfinnu}). The condition $|a|\ll\tilde
\Lambda_{\varepsilon}$ automatically leads to the inequalities $|a|\ll l_0$
and $|a|\ll\hbar/\sqrt{m\varepsilon}$. Hence, the function 
$w(\varepsilon/2\hbar\omega_0)$ is much smaller than $l_0/|a|$, unless
$\varepsilon\alt\hbar\omega_0\exp{(-l_0/|a|)}$ (see Eq.(\ref{wlog}) and
Fig.3). Accordingly, Eq.(\ref{eta}) gives $\eta=1$ and Eqs.~(\ref{Amplfin}),
(\ref{Amplfinnu}) lead to the scattering amplitudes
\begin{equation}   \label{conf3D}
f_{\nu\nu'}=4\pi a\varphi_{\nu}(0)\varphi_{\nu'}(0)\theta(\varepsilon-
\hbar\omega_0\nu)\theta(\varepsilon-\hbar\omega_0\nu').
\end{equation}   
The amplitudes (\ref{conf3D}) are nothing else than the 3D scattering
amplitude averaged over the axial distribution of particles in the incoming
($\nu$) and outgoing ($\nu'$) scattering channels. From Eqs.~(\ref{alphanu})
and (\ref{conf3D}) one obtains the same rate of transitions 
$\nu\rightarrow\nu'$ as in the case of 3D scattering of particles harmonically
confined  in the axial direction and interacting with each other via the
potential  $V(r)$. This is what one should expect, since under the condition 
$|a|\ll\tilde\Lambda_{\varepsilon}$ the amplitude of 3D scattering is momentum
independent. The integration over $d^2q$ in Eq.(\ref{baralpha}) leads to the
mean collisional frequency
\begin{eqnarray}    
\Omega=\frac{\hbar n}{2m}(4\pi a)^2A\sum_{\nu,\nu'}\varphi_{\nu}^2(0)
\varphi_{\nu'}^2(0)\exp{\left(-\frac{\hbar\omega_0}{T}{\rm max}
\{\nu,\nu'\}\right)}.    \nonumber
\end{eqnarray}

Thus, in the case where $|a|\ll\tilde\Lambda_T$, the tight 
confinement in the axial direction can manifest itself in the collisional 
rates only through the axial distribution of particles and  
the discrete structure of quantum levels in the axial confining potential.   
The expression for the collisional frequency $\Omega$ can be reduced to
the form
\begin{equation}        \label{Omegalowa}
\Omega=\frac{8\pi\hbar n}{m}\left(\frac{a}{l_0}\right)^2\xi,
\end{equation}
where the coefficient $\xi$ ranges from $1$ at $T\ll\hbar\omega_0$ to
$2/\pi$ for $T\gg\hbar\omega_0$. The condition $|a|\ll\tilde\Lambda_T$ is 
equivalent to $|a|\ll l_0$ and $q_T|a|\ll 1$, 
where $q_T=\sqrt{mT}/\hbar$ is the thermal
momentum of particles. For $T\gg\hbar\omega_0$, Eq.(\ref{Omegalowa}) gives
the collisional frequency which 
coincides with the three-dimensional result averaged over the classical 
Boltzmann profile of the 3D density in the axial direction, $n_B(z)$:
\begin{equation}       \label{Omega3D}
\Omega_{3D}=\langle\sigma_{3D} v\rangle\int\frac{ n^2_B(z)}{n}dz.
\end{equation}  
Here $\sigma$ is the 3D elastic cross section,  
and $v$ is the relative velocity of colliding particles. 
In other words, the quantity $(1/2)\bar\alpha n^2=(1/2)\Omega n$
coincides with the number of 3D collisions per unit time and unit surface
area in the $x,y$ plane, given by
$(1/2)\langle\sigma_{3D}v\rangle\int n_B^2(z)dz$.

From Eq.(\ref{Omegalowa}) we conclude that for $|a|\ll l_0$ the
confinement-dominated 3D regime of scattering is not pronounced.
At temperatures $T\alt\hbar\omega_0$ the collisional rate only slightly
deviates from the ordinary 3D behavior. This has a
simple physical explanation. For $|a|\ll\tilde\Lambda_T$, treating collisions
as three-dimensional we have $\Omega\sim 8\pi a^2vn_{3D}$. At low temperatures
$T\alt\hbar\omega_0$ the velocity $v\sim\hbar/ml_0$ and the
3D density is $n_{3D}\sim n/l_0$. For $T\gg\hbar\omega_0$ we
have $v\sim (T/m)^{1/2}$ and $n_{3D}\sim n(m\omega_0^2/T)^{1/2}$.
In both cases the "flux" $vn_{3D}\sim \omega_0 n$, and there is only a small
numerical difference between the low-T and high-T collisional frequencies.
  
The ultra-cold limit (\ref{u}) assumes that the characteristic radius 
of interatomic interaction $R_e\ll l_0$. Therefore, the condition
$|a|\ll l_0$ is always satisfied, unless the scattering length is
anomalously large ($|a|\gg R_e$). Below we will focus our attention
on this case, assuming that $|a|\agt l_0$. 

Let us first show how the 3D result follows from our analysis at 
$T\gg\hbar\omega_0$, irrespective of the relation between $a$ and 
$\tilde\Lambda_{\varepsilon}$. At these temperatures the main contribution 
to the sum in Eq.(\ref{baralpha})
comes from $\varepsilon\gg\hbar\omega_0$ and large $\nu$ and $\nu'$. 
Accordingly, we can replace the summation over $\nu$ and $\nu'$ by integration.
At energies much larger than $\hbar\omega_0$ the quantity $\sqrt{\nu}/l_0$ 
is nothing else than the axial momentum $k_z$ and we have 
$\varepsilon=\hbar^2(q^2+k_z^2)/m$.
For these energies the function $w(\varepsilon/2\hbar\omega_0)$ in 
Eq.(\ref{Amplfin}) takes its asymptotic form 
$w\approx i\sqrt{2\pi\varepsilon/\hbar\omega_0}$. Using Eq.(\ref{Amplfinnu}),
this immediately allows us to write 
$$|f_{\nu\nu'}|^2=P^2_{\nu\nu'}\frac{8\pi a^2}{l_0^2(1+p^2a^2)}=
P^2_{\nu\nu'}\frac{\sigma_{3D}}{l_0^2};\,\,\,\,\,\,
\nu'<\frac{\varepsilon}{\hbar\omega_0},$$
where $p=\sqrt{q^2+k_z^2}$ is the 3D momentum of the relative motion, and
$\sigma_{3D}=8\pi a^2/(1+p^2a^2)$ is the cross section for the 3D elastic 
scattering. For large $\nu$ and $\nu'$, Eq.({\ref{P}) gives
$P_{\nu\nu'}=(4/\pi^2\nu\nu')^{1/4}$, and the integration over $\nu'$ 
in Eq.(\ref{baralpha}) multiplies
$\sigma_{3D}$ by the relative 3D velocity $v$. Then, turning from the 
integration over $\nu$ to the integration over the axial momentum, we reduce 
Eq.(\ref{baralpha}) to
\begin{equation}    \label{Omega3DD}
\Omega=\int \frac{n\Lambda_T^2d^3p}{(2\pi)^3}(\sigma_{3D}v)
A\exp{\left (-\frac{\hbar^2p^2}{mT}\right) },
\end{equation}
and one can easily check that Eq.(\ref{Omega3DD}) coincides with the
three-dimensional result $\Omega_{3D}$ (\ref{Omega3D}).

In the limiting case, where the thermal momentum of particles
satisfies the inequality $q_T|a|\gg 1$, we obtain
\begin{equation}   \label{al}
\Omega_{3D}= \frac{16\hbar n}{m}\left(\frac{\hbar\omega_0}{T}
\right) ;\,\,\,\,\,\,\,q_T|a|\gg 1.
\end{equation}
In the opposite limit, where $q_T|a|\ll 1$, at temperatures 
$T\gg\hbar\omega_0$ we automatically have $|a|\ll\Lambda_T$ and, accordingly,
recover Eq.(\ref{Omegalowa}) with $\xi=2/\pi$:
\begin{equation}     \label{as}
\Omega_{3D}=\frac{16\hbar n}{m}\left(\frac{a}{l_0}\right)^2;\,\,\,\,\,\,\,
q_T|a|\ll 1.
\end{equation}

As mentioned in Section \ref{Sec.Quasi2D}, for $|a|\agt l_0$ 
the approximate border 
line between the quasi2D and confinement-dominated 3D regimes 
is $\varepsilon_*\approx 0.1\hbar\omega_0$. In the
 temperature
interval $\varepsilon_*<T<\hbar\omega_0$, 
the leading scattering channel will be the same as in the quasi2D case,
that is $\nu=\nu'=0$. However, the expression for the scattering 
amplitude $f_{00}$ is different. From Fig.3 and Eq.(\ref{Amplfin}) one 
concludes that the real part of the function $w$ 
can be neglected and the scattering amplitude takes the form
$$f_{00}=\frac{2\sqrt{2\pi}}{l_0/a+i\sqrt{\pi/2}}.$$
Then, retaining only the scattering channel $\nu=\nu'=0$, 
Eqs.~(\ref{alphanu}) and (\ref{baralpha}) yield
\begin{equation}   \label{Omegacd}
\Omega=\frac{8\pi\hbar n}{m}\left(\frac{a}{l_0}\right)^2
\frac{1-\exp{(-\hbar\omega_0/T)}}{1+\pi a^2/2l_0^2}.
\end{equation}  
The difference 
of Eq.(\ref{Omegacd}) from the quasi2D result of Eqs.~(\ref{f2}) and 
(\ref{alphaquasi}) is related to the absence of the logarithmic term 
in the denominator. This follows from the fact that now we omitted
the real part of the function $w$, which is logarithmically large in the
quasi2D regime. 

It is worth to note that for $l_0\gg |a|$, Eq.(\ref{Omegacd}) is only
slightly different from the 3D result (\ref{as}). This is consistent
with the above given analysis leading to Eq.(\ref{Omegalowa}). 

On the other hand, for large $|a|/l_0$ the difference between 
Eq.(\ref{Omegacd}) and the 3D result (\ref{al}) is significant. 
This originates from the fact that for a large scattering length 
$a$ the 3D amplitude of scattering in the ultra-cold limit depends 
on the particle momenta. 
For a tight axial confinement, treating collisions as 
three-dimensional, the relative momentum of colliding particles at temperatures 
$T\alt\hbar\omega_0$ is $\sim 1/l_0$ and it no longer depends on temperature.
Hence, the scattering rate is quite different from that in 3D.
Given these arguments, one expects a strongly pronounced confinement-dominated 
3D regime of scattering if the ratio $|a|/l_0\gg 1$. 

\begin{figure} 
\epsfxsize=\hsize 
\epsfbox{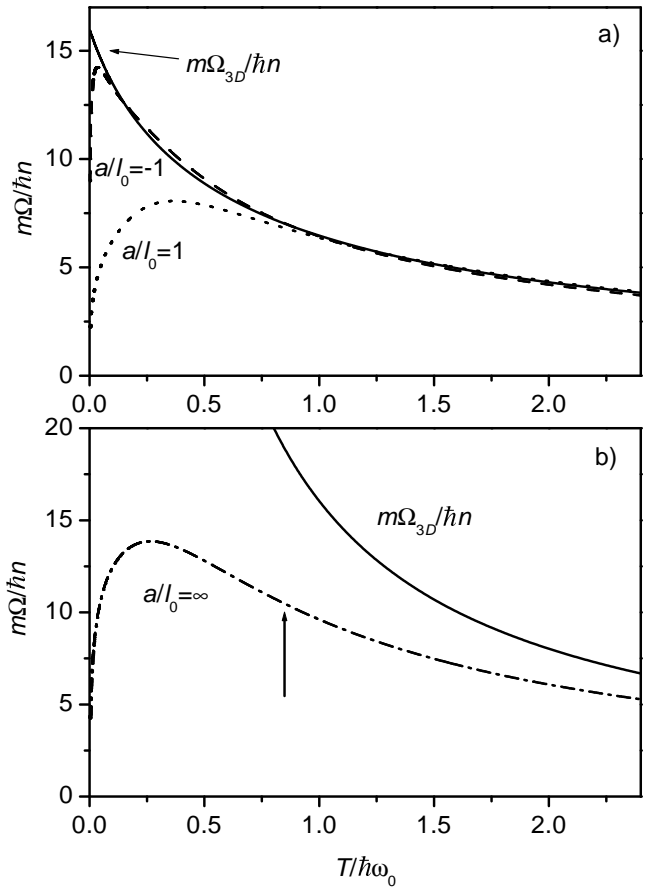} 
\caption{\protect
The quantity $m\Omega/\hbar n$ versus temperature. In (a) the parameter 
$a/l_0=-1$ (dashed curve), and $a/l_0=1$ (dotted curve). In (b) 
$a/l_0=\infty$ (unitarity limit). The solid curves in (a) and (b) show 
the 3D result (\ref{Omega3D}). The arrow in (b) indicates the lowest ratio 
$T/\hbar\omega_0$ in the Stanford and ENS experiments.}    
\label{5}   \end{figure}

This is confirmed by our numerical calculations for the temperature
dependence of $\Omega$ from Eq.(\ref{baralpha}). In Fig.5 we present the
results for $a/l_0$ equal to $-1$, $1$, and $\infty$. The largest deviation
from the 3D regime is observed in the unitarity limit ($a\rightarrow\infty$).
From Fig.5 we see that in the Stanford \cite{Chu2} and ENS \cite{Christ2} 
experiments performed in this limit \cite{comlength} one should have
significant  deviations of collisional rates from the ordinary 3D behavior. 

\section{Thermalization rates} \label{Sec.Therm.rates}

We will now discuss the collision-induced energy exchange between axial
and radial degrees of freedom of the particle motion in
an ultra-cold Bose gas tightly confined in the axial direction
of a pancake-shaped trap. It is assumed that the radial confinement is
shallow and it does not influence the scattering amplitudes. 
In this geometry, using degenerate Raman sideband cooling,  
the Stanford \cite{Chu1,Chu2} and ENS \cite{Christ1,Christ2} groups created 
Cs gas clouds with different axial ($T_z$) and radial ($T_{\rho}$)
temperatures. After switching off the cooling, interatomic collisions
lead to energy exchange between the axial and radial particle motion
and the temperatures $T_z$ and $T_{\rho}$ start to approach each other.
Ultimately, the gas reaches a new equilibrium state, with a
temperature in between the initial $T_z$ and $T_{\rho}$. The corresponding 
(thermalization) rate 
has been measured at Stanford \cite{Chu2} and ENS \cite{Christ2} and it
provides  us with the information on the regimes of interatomic collisions in
the gas. 

The radial motion of particles is classical. Therefore, we will calculate 
the rate of energy exchange between the radial and axial degrees of freedom 
for a given value of the radial coordinate \mbox{\boldmath$\rho$} and then 
average the result 
over the Boltzmann density profile in the radial direction. The latter is 
given by 
\begin{equation}     \label{Bprof}
n(\rho)=n(0)\exp{\left( -\frac{m\omega^2\rho^2}{2T}\right)},
\end{equation}
where $n(0)=m\omega^2N/2\pi T$ is the 2D density for $\rho=0$, $\omega$ the
radial frequency, and $N$ the total number of particles. Collision-induced
transitions $\nu\rightarrow\nu'$ change the energy of the 
axial motion by $\hbar\omega_0(\nu'-\nu)$. We will assume that in
the course of evolution the axial and radial distribution of particles remain 
Boltzmann, with instantaneous values of $T_z$ and $T_{\rho}$. Then the rate 
of energy transfer from the radial to axial motion can be written on the same
grounds as Eq.(\ref{baralpha}) and reads
\begin{eqnarray} \label{edot}
\dot E_z & = & -\dot E_{\rho}=\frac{1}{2}
\int n^2(\rho)d^2\rho\sum_{\nu\nu'}\int \frac{\Lambda_T^2d^2q}
{(2\pi )^2}\hbar\omega_0(\nu'-\nu)\times  \nonumber  \\
& & \frac{\hbar}{m}|f_{\nu\nu'}(\varepsilon)|^2 A
\exp{\left( -\frac{\hbar^2q^2}{mT_{\rho}}-\frac{\hbar\omega_0\nu}{T_z}\right)},
\end{eqnarray}   
where $\varepsilon=\hbar^2q^2/m+\hbar\omega_0\nu$, and the normalization 
coefficient $A$ depends now on both $T_z$ and $T_{\rho}$. 

The radial energy of the gas is $E_{\rho}=2NT_{\rho}$, and the axial energy
is given by $E_z=N\hbar\omega_0[\exp{(\hbar\omega_0/T)}-1]^{-1}$. The time 
derivatives of these energies take the form 
\begin{equation} \label{tz}
\dot{E}_z=\frac{N\hbar^2\omega^2_0 \dot{T}_z}{4T^2_z\sinh^2(\omega_0/2T_z)};
\,\,\,\,\,\,\dot{E}_{\rho}=2N \dot{T}_{\rho}.
\end{equation}
Given the initial values of $T_z$ and $T_{\rho}$, Eqs.(\ref{edot}) and 
(\ref{tz}) provide us with the necessary information on the
evolution of $T_z(t)$ and $T_x(t)$.

For a small difference $\delta T=T_{\rho}-T_z$, these equations can be 
linearized with respect to $\delta T$. As the total energy is conserved,
Eqs.(\ref{tz}) reduce to
\begin{equation}   \label{ddot}
\delta \dot{T}=\frac{\dot E_z}{N}\left(\frac{1}{2}+
\frac{4T^2\sinh^2(\hbar\omega_0/2T)}{\hbar^2\omega_0^2}\right).
\end{equation}  
In Eq.(\ref{edot}) we represent the exponent as 
$-(\varepsilon/T+\delta T\hbar\omega_0\nu/T^2)$ and turn from the integration
over $dq$ to integration over $d\varepsilon$. The zero
order term of the expansion  in powers of $\delta T$ vanishes. The (leading)
linear term, being substituted into Eq.(\ref{ddot}), leads to the differential
equation for  $\delta T(t)$:
\begin{equation} \label{diff}
\delta \dot{T}=-\Omega_{th}(T)\delta T,
\end{equation}
where the thermalization rate $\Omega_{th}(T)$ is given by
\begin{eqnarray} \label{rate}
\Omega_{th} & = & \frac{n(0)\Lambda_T^2A}{16\pi\hbar}
\left(\frac{\hbar^2\omega^2_0}{2T^2}+4\sinh^2(\hbar\omega_0/2T)\right)
\times  \nonumber  \\
& & \sum_{\nu>\nu'}(\nu-\nu')^2\int_{\hbar\omega_0\nu}^{\infty}d\varepsilon
|f_{\nu\nu'}(\varepsilon)|^2\exp\left(-\frac{\varepsilon}{T}\right). 
\end{eqnarray} 
The degeneracy parameter is $n(0)\Lambda_T^2=N(\hbar\omega/T)^2$ and 
it is small as
the gas obeys the Boltzmann statistics. The normalization coefficient
$A$ is again given by Eq.(\ref{A}).

The quantum numbers $\nu$ and $\nu'$ take only even values and, hence,
in order to change the state of the axial motion one should have a relative
energy $\varepsilon>2\hbar\omega_0$. Therefore, at temperatures 
lower than $\hbar\omega_0$ the rate of transitions changing the axial and
radial energy is $\propto\exp{(-2\hbar\omega_0/T)}$. On the other
hand, the axial energy $E_z\propto\exp{(-\hbar\omega_0/T)}$ and thus the 
thermalization rate $\Omega_{th}\propto\exp{(-\hbar\omega_0/T)}$. This can
be easily found from Eq.(\ref{rate}) and shows that the quantum
character of the axially confined particle motion exponentially suppresses
the thermalization process at temperatures $T\ll\hbar\omega_0$. In particular, 
this is the case for the quasi2D regime. 

In the most interesting part of the confinement-dominated 3D regime, where 
$\varepsilon_*<T<\hbar\omega_0$, the energy exchange between the axial and 
radial motion of particles is mostly related to transitions between the 
states with $\nu'=0$ and $\nu=2$. The relative energy $\varepsilon$ should
be larger than $2\hbar\omega_0$ and, at the same time, this energy is well
below $4\hbar\omega_0$. Hence, the scattering amplitude $f_{20}(\varepsilon)$
is determined by Eqs.~(\ref{Amplfin} and (\ref{Amplfinnu}) in which one can
put $w(\varepsilon/2\hbar\omega_0)\approx i3\pi/2$ (see Eq.(\ref{Im}) and 
Fig.3). This gives
$$f_{20}=\frac{2\sqrt{2\pi}}{l_0/a+i(3/2)\sqrt{\pi/2}},$$ 
and from Eq.(\ref{rate}) we obtain
\begin{equation} \label{ratelowtemp}
\!\!\Omega_{th}\!=\!\frac{16\hbar\omega_0}{9\pi T}\Omega_0
\frac{\exp(-\hbar\omega_0/T)}{1+8l_0^2/9\pi a^2}(1-\exp{(-\hbar\omega_0/T)})^3.
\end{equation}
The characteristic frequency $\Omega_0$ is given by
\begin{equation}    \label{omegam}
\Omega_0=\omega^2N/\omega_0.
\end{equation}

At temperatures $T\gg\hbar\omega_0$, Eq.(\ref{rate}) leads to the 3D result 
for the thermalization rate:
\begin{eqnarray}  
\Omega^{3D}_{th}&=&\frac{8}{15\pi}\left(\frac{\hbar\omega_0}{T}\right)^2
\Omega_0;\,\,\,\,\,\,\,q_T|a|\gg 1, \label{first}\\
\Omega^{3D}_{th}&=&\frac{16}{15\pi}\left(\frac{a}{l_0}\right)^2
\left(\frac{\hbar\omega_0}{T}\right)\Omega_0;
\,\,\,\,\,\,\,\,q_T|a|\ll 1.   \label{second}
\end{eqnarray}
Comparing Eqs.~(\ref{first}) and (\ref{second}) with Eq.(\ref{ratelowtemp}) 
one sees that $\Omega_{th}$ should acquire its maximum value at 
$T\sim\hbar\omega_0$. For $|a|\agt l_0$ this maximum value is on the order of
$\Omega_0/2\pi$. 

As one expects from the discussion in Section \ref{Sec.Conf-dom}, 
the difference of the thermalization rate from $\Omega_{th}^{3D}$ is 
pronounced for large values of $a$. For example, in the 
unitarity limit Eq.(\ref{first}) gives $\Omega^{3D}_{th}\propto 1/T^2$,
whereas in the confinement-dominated regime we have 
$\Omega_{th}\propto (1/T)\exp{(-\hbar\omega_0/T)}$.  

It should be emphasized that for any $T$, $\omega_0$, and $a$ the ratio
$\Omega_{th}/\Omega_0$ depends only on the parameters $T/\hbar\omega_0$
and $a/l_0$. This can be found directly from Eq.(\ref{rate}).
In Fig.6 we present the temperature dependence of $\Omega_{th}$ in
the unitarity limit, obtained numerically from Eq.(\ref{rate}), and compare
our results with the data of the Stanford \cite{Chu2} and ENS \cite{Christ2}
experiments. With the current error bars, the ENS results do not
show significant deviations from the classical 3D behavior. These results 
agree fairly well with our
 calculations. The Stanford experiment gives
somewhat lower values of
 $\Omega_{th}/\Omega_0$ at the lowest temperatures 
of the
 experiment. 
 
\begin{figure} 
\epsfxsize=\hsize 
\epsfbox{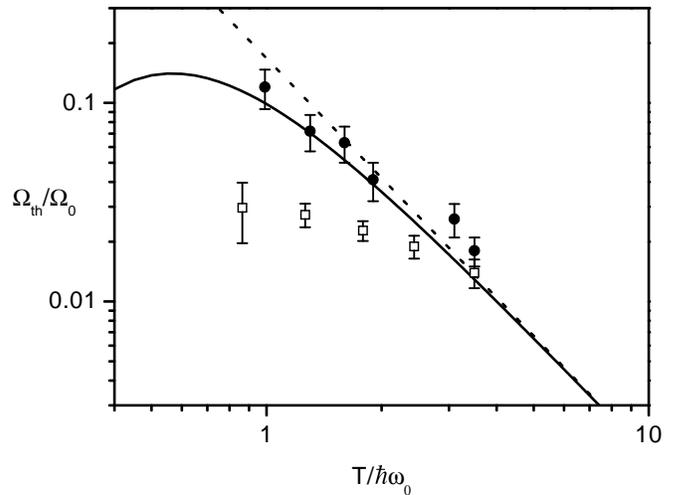} 
\caption{\protect
Thermalization rate versus temperature in the unitarity limit ($a=\infty$).
The solid curve shows the result of our calculations for
$\Omega_{th}/\Omega_0$, and the dotted line the 3D result
$\Omega_{th}^{3D}/\Omega_0$. Squares and circles show the 
data of the Stanford and ENS experiments.}   
\label{6}  
\end{figure}

In the hydrodynamic regime for the gas cloud, where the characteristic
collisional frequency greatly exceeds the radial frequency $\omega$, 
our assumption of quasiequilibrium at instantaneous (time-dependent) values
of $T_z$ and $T_{\rho}$ may not be valid. Nevertheless, the shape of the
curve $\Omega_{th}(T)$ qualitatively remains the same, including the
exponential decrease with temperature at $T<\hbar\omega_0$ and power law
decrease with increasing $T$ at temperatures larger than $\hbar\omega_0$.
However, the maximum value of $\Omega_{th}$ will be somewhat lower
(in particular, of the order of $\omega$ \cite{Chu2}). 

The number of particles (per ``2D'' sheet of atoms) in the Stanford 
experiment \cite{Chu2} was $N\sim 10^4$ \cite{private}, which is by 
a factor of 20 higher than at ENS for $T\approx\hbar\omega_0$ \cite{Christ2}.
We then estimate the 2D density of atoms for these temperatures to be
$n\sim 2.5\times 10^8$ cm$^{-2}$ at Stanford ($\omega\approx 90$ Hz), 
and $n\sim 0.5\times 10^8$ cm$^{-2}$ at ENS ($\omega\approx 180$ Hz). 
For these densities, the ratio of the collisional frequency $\Omega$ in
Fig.5 to the radial frequency is $\Omega/\omega\sim 0.3$ in the ENS
experiment, and $\Omega/\omega\sim 3$ in the experiment at Stanford.
At temperatures $T>\hbar\omega_0$ the density $n$ and the ratio 
$\Omega/\omega$ are smaller in both experiments.
We thus see that the ENS experiment \cite{Christ2} was in the 
collisionless regime, although rather close to the hydrodynamic regime 
at temperatures $T\approx\hbar\omega_0$. For these temperatures, the 
Stanford experiment \cite{Chu2} has already achieved the hydrodynamic 
regime, and this can explain the discrepancy between our calculations and 
the Stanford results in Fig.6.

\section{Inelastic 2-body processes} \label{Sec.Inelastic}

Inelastic scattering of atoms is also influenced by the tight axial confinement
of the particle motion. In this Section we will consider the inelastic
2-body processes, such as spin relaxation, in which the internal states of
colliding atoms are changing, and the released internal-state energy of the
atoms is transferred to their kinetic energy. Our goal is to establish a 
relation between the inelastic rates in 3D and those in the (tightly)
axially confined geometry. The analysis given below relies on two important
conditions widely met for the 2-body spin relaxation \cite{paul}:\\
i)The energy release per collision greatly exceeds the gas temperature
and the frequency of the axial confinement. Accordingly, the inelastic
transitions occur at comparatively short interparticle distances $\sim R_{in}$
which are much smaller than the characteristic de Broglie wavelength of 
particles.\\
ii)The inelastic transitions are caused by weak (spin-dipole, spin-orbit, etc.)
interatomic interactions and can be treated with perturbation 
 theory.

To first order in perturbation theory the amplitude of inelastic scattering,
defined in the same way as in the previous Sections, is given by a general
expression \cite{Dav}
\begin{equation}    \label{inf}
f_{in}(\varepsilon)=\frac{m}{\hbar^2}\int\psi_i({\bf r})U_{int}({\bf r})
\psi_f({\bf r})d^3r.
\end{equation} 
Here $\psi_i({\bf r})$ and $\psi_f({\bf r})$ are the true wavefunctions of the
initial and final states of the relative motion of colliding atoms, and 
$U_{int}({\bf r})$ is the (weak) interatomic potential responsible for 
inelastic transitions. This potential is the same as in the 3D case. 
The function $\psi_f$ is also the same as in 3D, since the relative
energy in the final state is much larger than $\hbar\omega_0$. 
Thus, the only difference of the amplitude $f_{in}$ (\ref{inf}) from
the amplitude of inelastic scattering in the 3D case is related
to the form of the wavefunction $\psi_i$. 

The characteristic interatomic distance $R_{in}$ at which the inelastic 
transitions occur, satisfies the inequality 
$R_{in}\ll\tilde\Lambda_{\varepsilon}$
(see item ii). Therefore, we are in the ultra-cold limit similar to that 
determined by Eq.(\ref{u}) in the case of elastic scattering, and the 
conditions  $qR_{in}\ll 1$ and $R_{in}\ll l_0$ are satisfied. The former
ensures a dominant contribution of the $s$-wave (of the initial wavefunction
$\psi_i$) to the scattering amplitude $f_{in}$ (\ref{inf}). Due to the
condition $R_{in}\ll l_0$, at distances $r\sim R_{in}$ the
wavefunction $\psi_i$ has a three-dimensional character:
$\psi_i(r)\propto\tilde\psi_{3D}(r)$, where $\tilde\psi_{3D}(r)$ is the
wavefunction of the 3D relative motion at zero energy. For $r\gg R_e$ we have
$\tilde\psi_{3D}(r)=(1-a/r)$, and in order to be consistent with Eq.(\ref{3D})
we should write  
\begin{equation}    \label{psii}
\psi_i(r)=\eta(\varepsilon)\varphi_{\nu}(0)\tilde\psi_{3D}(r), 
\end{equation}
where the coefficient $\eta(\varepsilon)$ is given by Eq.(\ref{eta}), and 
$\nu$ is the quantum number of the initial state of the relative motion 
in the axial harmonic potential $V_H(z)$.

In the 3D case, the amplitude $f_0$ of inelastic scattering at zero initial
energy is determined by Eq.(\ref{inf}) with $\psi_i$ replaced
by $\tilde\psi_{3D}$. Hence, Eq.(\ref{psii}) directly gives a relation 
between the two scattering amplitudes: 
\begin{equation}    \label{relation}
f_{in}=\eta(\varepsilon)\varphi_{\nu}(0)f_0.
\end{equation}
Due to the high relative kinetic energy of particles in the final state of the
inelastic channel, the density of final states in this channel is
independent of the axial confinement. Therefore, relying on 
Eq.(\ref{relation}) the mean rate 
 constant $\bar\alpha_{in}$ of inelastic
collisions in the axially confined
 geometry and the corresponding collisional
frequency can be represented 
 in the form
\begin{equation}    \label{rcin}
\bar\alpha_{in}=\langle|\eta(\varepsilon)|^2\varphi_{\nu}^2(0)
\theta(\varepsilon-\hbar\omega_0\nu)\rangle\alpha_0;
\,\,\,\,\,\,\Omega_{in}=\bar\alpha_{in}n,
\end{equation} 
where $\alpha_0$ is the 3D inelastic rate constant at zero energy.

Note that in the ultra-cold limit the 3D inelastic rate constant is temperature
independent and equal to $\alpha_0$ if the scattering length $|a|\alt R_e$.
For $|a|\gg R_e$, the wavefunction of the relative motion in the region of
interatomic interaction takes the form $\psi_i(r)=\eta_{3D}\tilde\psi_{3D}(r)$,
where $|\eta_{3D}|^2=(1+p^2a^2)^{-1}$ and $p$ is the 3D relative momentum 
of colliding particles (see, e.g. \cite{FRRS}). Hence,  
for the inelastic rate constant we have 
$\langle\alpha_0(1+p^2a^2)^{-1}\rangle$. In the presence of axial confinement,
averaging the frequency of inelastic collisions over the (quantum) axial 
density profile $n_{3D}(z)$, we obtain 
\begin{equation}     \label{Omegain3D}
\Omega_{in}=\left\langle\frac{\alpha_0}{(1+p^2a^2)}\right\rangle
\int\frac{n_{3D}^2(z)}{n}dz.
\end{equation} 
The density profile $n_{3D}(z)$ accounts for the discrete structure of
quantum levels in the axial confining potential and for the quantum spatial
distribution of particles. Therefore, Eq.(\ref{Omegain3D}) gives the 
ordinary 3D result only at temperatures $T\gg\hbar\omega_0$, where $n_{3D}(z)$
becomes the Boltzmann distribution $n_B(z)$. 

We first analyze the influence of axial confinement on $\Omega_{in}$ 
(\ref{rcin}) for the case where $|a|\ll\tilde\Lambda_T$ or, equivalently, 
$|a|\ll l_0$ and $q_T|a|\ll 1$. In this case we may put $\eta=1$ at any $T$, 
except for extremely low temperatures in the quasi2D regime. Then
Eq.(\ref{rcin})  gives 
\begin{equation}     \label{ainav3D}
\Omega_{in}=\langle\varphi_{\nu}^2(0)\rangle\alpha_0n=
\frac{\alpha_0n}{\sqrt{2\pi}\,l_0}\tanh^{1/2}{\left(
\frac{\hbar\omega_0}{T}\right)}.
\end{equation}
One can easily check that Eq.(\ref{ainav3D}) coincides with 
Eq.(\ref{Omegain3D}) in which $p|a|\ll 1$.
The reason for this coincidence is that, similarly to the case of elastic 
scattering described by Eq.(\ref{conf3D}), for $\eta=1$ the scattering
amplitude $f_{in}$ (\ref{relation}) is independent of the relative energy 
$\varepsilon$. Hence, the inelastic rate is influenced by the axial
confinement only through the axial distribution of particles.
However, this influence is significant, in contrast to the case of elastic
scattering under the same conditions (see Eq.(\ref{Omegalowa})). 
Qualitatively, for $q_T|a|\ll 1$
we have $\Omega_{in}\sim\alpha_0n_{3D}$. At temperatures $T\ll\hbar\omega_0$,
a characteristic value of the 3D density is $n_{3D}\sim n/l_0$ and we obtain
$\Omega_{in}\sim\alpha_0n/l_0$. For $T\gg\hbar\omega_0$, the 3D density
$n_{3D}\sim n (m\omega_0^2/T)^{1/2}$ and hence the frequency of inelastic
collisions is $\Omega_{in}\sim (\alpha_0n/l_0)(\hbar\omega_0/T)^{1/2}$. 
 
We now discuss the temperature dependence of the inelastic rate for 
the case where $|a|\agt l_0$, which in the ultra-cold limit (\ref{u})
assumes that $|a|\gg R_e$. In the quasi2D regime and in the temperature 
interval $\varepsilon_*<T<\hbar\omega_0$
of the confinement-dominated 3D regime, the most important contribution to
the inelastic rate in Eq.(\ref{rcin}) comes from collisions with the axial
quantum number $\nu=0$. Then, using Eq.(\ref{Amplfin}) we express the
parameter  $\eta$ through the elastic amplitude $f_{00}$ and obtain a relation
between $\Omega_{in}$ and the mean frequency of elastic collisions $\Omega(T)$:
\begin{equation}     \label{ainlow}
\Omega_{in}(T)=\langle|f_{00}(\varepsilon)|^2
\rangle\frac{\alpha_0n}{(4\pi a\varphi_0(0))^2}=\Omega(T)\beta,
\end{equation}
where $\beta=(1/128\pi^3)^{1/2}\!(ml_0\alpha_0/\hbar a^2)$ is a dimensionless
parameter independent of temperature. The temperature dependence of $\Omega$
is displayed in Fig.1 and Fig.5 and was discussed in sections 
\ref{Sec.Quasi2D} and \ref{Sec.Conf-dom}. Note that the parameter $\beta$
is not equal to zero for $|a|\rightarrow\infty$. In this case, since the
amplitude $f_0$ is  calculated with the wavefunction $\tilde\psi_{3D}$
which behaves as $a/r$ for $r\rightarrow\infty$, we have $\alpha_0
\propto a^2$ and $\beta={\rm const}$.   

At temperatures $T\gg\hbar\omega_0$, using the same method as in Section
\ref{Sec.Conf-dom} for the case of elastic scattering, from Eq.(\ref{rcin}) we
recover the 3D result $\Omega_{in}^{3D}$ given by Eq.(\ref{Omegain3D}) with 
$n_{3D}(z)=n_B(z)$. In the limiting case, where $q_T|a|\gg 1$, we find
\begin{equation}   \label{in3}
\Omega_{in}^{3D}=\frac{8\hbar n}{m}\beta\left(\frac{2\hbar\omega_0}{T}
\right)^{3/2}=\Omega_{3D}(T)\beta\left(\frac{2\hbar\omega_0}{T}\right)^{1/2}.
\end{equation}
Comparing Eq.(\ref{in3}) with Eq.(\ref{ainlow}), we see that in the 
confinement-dominated 3D regime the deviation of the inelastic rate from
the ordinary 3D behavior should be larger than that in the case of elastic 
scattering.

As follows from Eq.(\ref{ainlow}) and Fig.5, for $|a|\agt l_0$ the inelastic
frequency $\Omega_{in}$ reaches its maximum at temperatures near the border
between the quasi2D and confinement-dominated 3D regimes. The maximum value
of $\Omega_{in}$ is close to
\begin{equation}    \label{Omegaintilde}
\tilde\Omega_{in}=\frac{16\hbar}{m}\beta.
\end{equation}
From Eq.(\ref{rcin}) one finds that at any $T$ the ratio
$\Omega_{in}/\tilde\Omega_{in}$ depends only on two parameters:
$T/\hbar\omega_0$ and $a/l_0$. In Fig.7 we present our numerical results for
$\Omega_{in}/\tilde\Omega_{in}$ as a function  of $T/\hbar\omega_0$ for
$a/l_0$ equal to $-1$, $1$, and $\infty$. As expected, the deviations from the
3D behavior are the largest in the unitarity limit.   

\begin{figure} 
\epsfxsize=\hsize 
\epsfbox{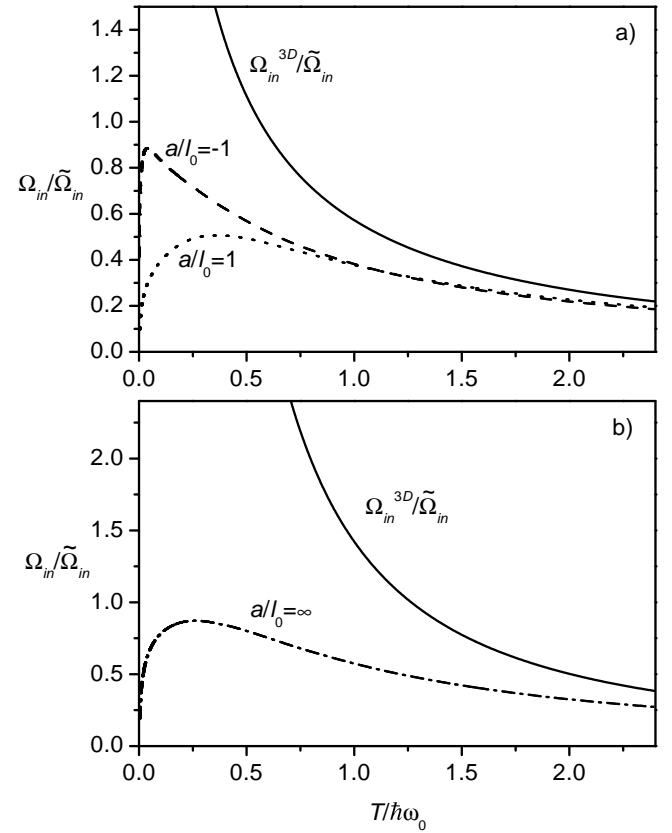} 
\caption{\protect
The quantity $\Omega_{in}/\tilde\Omega_{in}$ versus temperature. In (a) the parameter 
$a/l_0=-1$ (dashed curve), and $a/l_0=1$ (dotted curve). In (b) 
$a/l_0=\infty$ (unitarity limit). The solid curves in (a) and (b) show 
the 3D result (\ref{ainlow}).}    
\label{7}   \end{figure}

The inelastic rate of
spin relaxation in a tightly (axially) confined gas of atomic cesium has been
measured for the unitarity limit in the  Stanford experiment \cite{Chu2}.
Due to a shallow radial confinement of the cloud in this experiment, 
the 2D density $n\sim 1/T$ (see Eq.(\ref{Bprof})). Then, 
Eq.(\ref{in3}) gives the 3D inelastic frequency 
$\Omega^{3D}_{in}\sim 1/T^{5/2}$, whereas in the temperature interval 
$\varepsilon_*<T<\hbar\omega_0$ of the confinement-dominated regime
Eqs.~(\ref{Omegacd}) and (\ref{ainlow}) lead to 
$\Omega_{in}\sim (1/T)(1-\exp{(-\hbar\omega_0/T)})$.
In order to compare our calculations with the data of the Stanford
experiment on spin relaxation, in Fig.8 we display the ratio of 
$\Omega_{in}(T/\hbar\omega_0)$
to $\Omega_{in}$ at $T=3\hbar\omega_0$ which was the highest temperature
in the experiment. The temperature dependence of the inelastic rate, following
from the Stanford results, agrees fairly well with the calculations and
shows significant deviations from the 3D behavior. It should be noted that,
in contrast to thermalization rates, the inelastic decay rate is not
sensitive to whether the gas is in the collisionless or hydrodynamic regime
\cite{Chu2}.
\begin{figure} 
\epsfxsize=\hsize 
\epsfbox{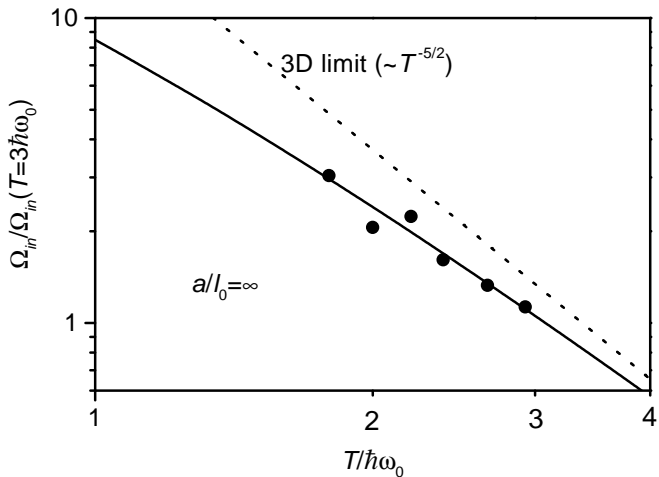} 
\caption{\protect
The ratio of $\Omega_{in}(T/\hbar\omega_0)$ to $\Omega_{in}$ at
$T=3\hbar\omega_0$ versus temperature in the unitarity limit ($a=\infty$).
The solid curve shows the result of our numerical calculations, and the dotted
line the 3D limit. Circles show the data of the Stanford experiment.}     
\label{8}  
\end{figure}

\section{Concluding remarks}    \label{Sec.Concl}

In conclusion, we have developed a theory which describes the influence
of a tight axial confinement of the particle motion on the processes of 
elastic
and inelastic scattering. The most interesting case turns out to be the one
in which the 3D scattering length $a$ exceeds the extension of the 
wavefunction in the axial 
direction, $l_0$. In the ultra-cold limit defined by Eq.(\ref{u}), the
condition $|a|>l_0$
automatically requires large $|a|$ compared to the radius of interatomic
interaction $R_e$. Then we have a pronounced confinement-dominated 3D regime
of scattering at temperatures on the order of $\hbar\omega_0$. Treating
interatomic
 collisions as three-dimensional, the relative momentum of
colliding atoms
 is related to the quantum character of the axial motion in
the confining
 potential and becomes of the order of $1/l_0$. As a result, the
scattering rate
 can strongly deviate from the ordinary 3D behavior. The axial
extension of the
 wavefunction, achieved in the experiments at Stanford and
ENS
 \cite{Chu1,Chu2,Christ1,Christ2}, is $l_0\approx 200$ \AA. The required
value
 of the scattering length, $|a|>l_0$ and $|a|\gg R_e$, is characteristic
for Cs
 atoms ($R_e\approx 100$\AA$\,$) and  can also be achieved for other
alkali
 atoms by using Feshbach resonances. 

In order to observe the 2D features of the particle motion in the rates
of interatomic collisions one has to reach the quasi2D regime of scattering,
which requires much lower temperatures, at least by an order of magnitude 
smaller than $\hbar\omega_0$. For $\omega_0\approx 80$kHz ($\hbar\omega_0
\approx 4$ $\mu$K) as in the Stanford \cite{Chu1,Chu2} and ENS
\cite{Christ1,Christ2} experiments, these are temperatures below $400$nK. As
one can see from Fig.5, the rate of  elastic collisions is still rather large
for these temperatures and, hence, one can think of achieving them by
evaporative cooling. Moreover, for realistic radial  frequencies $\omega\sim
100$Hz there is a hope to achieve quantum degeneracy and observe a cross-over
to the BEC regime. The cross-over temperature is $T_c\approx
N^{1/2}\hbar\omega$ (see \cite{Kleppner} and the discussion in \cite{PHS}),
and for $N\sim 1000$ particles in a quasi2D layer we find $T_c\approx 100$nK.

Another approach to reach BEC in the quasi2D regime will be to prepare
initially a 3D trapped condensate and then adiabatically slowly turn
on the tight axial confinement. Manipulating the obtained (quasi)2D
condensate and inducing the appearance of thermal clouds with temperatures
$T<T_c$, one can observe interesting phase coherence phenomena originating 
from the phase fluctuations of the condensate in quasi2D (see \cite{PHS}). 

Interestingly, at temperatures $T\sim T_c$ the collisional frequency 
$\Omega$ can be on the order of the cross-over temperature if $|a|\agt l_0$. 
This follows directly from Fig.1 and Eq.(\ref{alphaquasi}) which give 
$\Omega\sim \pi\hbar n/m$ even at temperatures by two orders of magnitude 
smaller than $\hbar\omega_0$. As the 2D density of thermal particles is 
$n\sim Nm\omega^2/T$, we immediately obtain $\hbar\Omega(T_c)\sim
N^{1/2}\hbar\omega\approx T_c$. This condition means that the trapped gas
becomes a strongly interacting system. The mean free path of a particle,
$v/\Omega(T_c)$, is already on the order of its de Broglie wavelength
$\hbar/\sqrt{mT_c}$. At the same time, the system remains dilute, since the 
mean interparticle separation is still much larger than the radius of 
interatomic interaction $R_e$. In this respect, the situation is similar 
to the 3D case
with a large scattering length $a\gg R_e$ at densities where $na^3\sim 1$.
The investigation of the cross-over to the BEC regime in such strongly 
interacting quasi2D gases should bring in analogies with condensed matter 
systems or dense 2D gases. Well-known examples of dense 2D systems in which 
the Kosterlitz-Thouless phase transition \cite{KTT} has
been found experimentally, are monolayers of liquid helium \cite{KTTexp}     
and the quasi2D gas of atomic hydrogen on liquid helium surface
\cite{Simo}.
       
On the other hand, for $|a|\ll l_0$ the collisional frequency near the
BEC cross-over, $\Omega(T_c)\ll T_c/\hbar$, and the (quasi)2D gas remains
weakly interacting. Then, the nature of the cross-over is questionable
(see \cite{PHS}). Generally speaking, one can have both the ordinary
BEC cross-over like in an ideal trapped gas \cite{Kleppner} and the 
Kosterlitz-Thouless type \cite{KTT} of a cross-over. We thus see that 
axially confined Bose gases in the quasi2D regime are remarkable systems 
where by tuning $a$ or $l_0$ one can modify the nature of the BEC cross-over.

\section*{Acknowledgements}

We acknowledge fruitful discussions with I. Bouchoule, C. Salomon, A.J.
Kerman, V. Vuleti\'c, and J.T.M. Walraven. This work was financially supported
by the Stichting voor Fundamenteel Onderzoek der Materie (FOM),
by INTAS, and by the Russian Foundation for Basic Studies. 

%In the unitarity limit ($a\rightarrow \infty$) the
%temperature dependence of the inelastic decay rate (proportional to $\nu$) 
%was observed in
%\cite{Chu}. In this limit the step-like structure in Fig.(1) is even more
%pronounced and one should clearly see $T^{-1}$ dependence of the rate for small
%temperatures. In the experiment of Chu the desired limit $T\ll \omega_0$ was
%not achieved.


\begin{references}
\bibitem{com3} For interatomic potentials which have a shallow well
(spin-polarized atomic hydrogen), the bound state with an exponentially small
binding energy $\varepsilon_0$, characteristic for shallow 2D potential wells
\cite{LLQ}, does not exist due to the presence of a strong repulsive core
\cite{Kagan2D}. For potentials with a deep well (alkali atoms) the situation
is similar to that in 3D: There are many bound states and only accidentally
one can encounter a very weakly bound $s$-level ($d_*\rightarrow\infty$). Thus,
for realistic momenta of particles in ultracold gases we always have
the inequality $qd_*\ll 1$.
\bibitem{com1} This is easily established by comparing the relation for the
vertex of elastic interaction, obtained to zero order in perturbation theory
\cite{AGD}, with the expression for the scattering amplitude \cite{LLQ}.
In the case of an inhomogeneous density profile one should replace
$\mu$ by $n_0g$, where $n_0$ is the coordinate-dependent condensate density.
\bibitem{com2} In \cite{PHS} the coefficient $B$ was put equal to unity.
\bibitem{comlength} In the Stanford experiment \cite{Chu2} the unitarity limit
for collisions between Cs atoms in the hyperfine state $F=3$, $m_F=3$ 
has been achieved by tuning the magnetic field to the Feshbach
resonance at 30 G. In the ENS experiment \cite{Christ2} the magnetic field
was very small. According to the recent calculations \cite{Leo}, this
corresponds to a large and negative scattering length ($a\sim -1500$ \AA$\,$)
for collisions between Cs atoms in $F=3$ states. For the ENS axial frequency
$\omega_0=80$ kHz we have $l_0\approx 200$ \AA$\,$. Hence, the ratio
$|a|/l_0$ in this experiment was approximately equal to 7, which is already
rather close to the unitarity limit.
\end{references}
\end{document}